\documentclass[twocolumn,english,american,aps,amsfonts,amsmath,prd,tightenlines,nofootinbib]
{revtex4}
\usepackage[T1]{fontenc}
\usepackage[latin1]{inputenc}
\usepackage{amssymb}
\usepackage{amsmath}

\makeatletter

\usepackage{epsf}

\newcommand{\beq}{\begin{equation}}
\newcommand{\eeq}{\end{equation}}
\newcommand{\beqn}{\begin{eqnarray}}
\newcommand{\eeqn}{\end{eqnarray}}

\newcommand{\Tr}{{\rm Tr}}

\newcommand{\be}{\begin{equation}}
\newcommand{\ee}{\end{equation}}
\newcommand{\ba}{\begin{eqnarray}}
\newcommand{\ea}{\end{eqnarray}}
\newcommand{\bdm}{\begin{displaymath}}
\newcommand{\edm}{\end{displaymath}}

\newcommand{\ie}{{\it i.e.\ }}

\DeclareMathAlphabet{\mathpzc}{OT1}{pzc}{m}{it}
\def\bea{\begin{eqnarray}}
\def\eea{\end{eqnarray}}
\def\beas{\begin{eqnarray*}}
\def\eeas{\end{eqnarray*}}
\def\sla{\raise.15ex\hbox{$/$}\kern-.57em}

\def\bea{\begin{eqnarray}}
\def\eea{\end{eqnarray}}

\def\sla{\raise.15ex\hbox{$/$}\kern-.57em}
\def\ie{{\it i.e.}~}

\def\cF{{\cal F}}

\def\cH{{\cal H}}
\def\cI{{\cal I}}

\def\cN{{\cal N}}

\def\cQ{{\cal Q}}

\def\cV{{\cal V}}

\begin{document}

%\begin{flushright}
%\noindent {ROM2F/2009/21},
%{CERN-PH-TH/2009-198}, \\
%{SU-ITP/2009/45}
%\end{flushright}

\title{\Large\bf Perturbative and Non-perturbative $\cN =8$ supergravity}

%\title{\Large\bf
%$\cN =8$ Supergravity, UV finiteness and Cosmic Censorship}

\author{{\bf Massimo Bianchi$^{1,2}$}, {\bf Sergio Ferrara$^{1,3,4}$},
and {\bf Renata Kallosh$^5$} }

%\date{\today}

\affiliation{ $^1${\sl Physics Department, Theory Unit, CERN  CH
1211, Geneva 23, Switzerland }$^2${\sl Dipartimento di Fisica and
Sezione INFN, Universit\`a di Roma ``Tor Vergata'', Via della
Ricerca Scientifica, 00133 Roma, Italy} $^3${\sl INFN - Laboratori
Nazionali di Frascati,  Via Enrico Fermi 40, 00044 Frascati,
Italy} $^4${\sl Department of Physics and Astronomy University of
California, Los Angeles, CA USA} $^5${\sl Department of Physics,
Stanford University, Stanford, CA 94305}}

%\selectlanguage{american}
\begin{abstract}
We study extremal black holes, their ADM mass and area of the
horizon in $\cN = 8$ supergravity. Contrary to intuition gained
from $\cN = 2,3, 4$  theories, in $\cN = 8$, as well as in
$\cN=5,6$, supergravity BPS states may become massless only at the
boundary of moduli space. We show that stringy states described in
\cite{Green:2007zzb}, which have no mass gap and survive in
toroidal compactifications in addition to the massless states of
perturbative $\cN = 8$ supergravity, display a null singularity in
four-dimensional space-time, when viewed as solutions of the $\cN
= 8$ version of Einstein equations.  We analyze known methods of
resolving such singularities and explain why they all fail in
$D=4$,  $\cN = 8$ supergravity. We discuss possible implications
for the issue of UV finiteness of perturbative $\cN = 8$
supergravity  and the plausibility of a non-perturbative
completion that exclude singular states.

\end{abstract}

\ \maketitle {\it Introduction} -- $\cN =8$ supergravity
\cite{Cremmer:1979up} is the theory with the largest possible
amount of supersymmetry for particles with spin $s\le 2$ in $D=4$.
The perturbative theory has recently shown surprisingly  good UV
properties \cite{Bern:2009kd}. In particular, at three loops the
expected UV divergent $R^4$ term \cite{Kallosh:1980fi}  is absent,
which could have broken the continuous $E_{7(7)}$  symmetry to a
discrete subgroup \cite{Hull:1994ys} in combination with instanton
effects. Not even at four loops the theory produces any divergence
\cite{Bern:2009kd}. Similarly to $\cN=5,6$ supergravities and
contrary to other extended supergravity theories with $\cN<5$,
$E_{7(7)}$ symmetry is not anomalous \cite{Marcus:1985yy}. As for
the $SL(2,R$ symmetry of Type IIB superstrings in $D=10$
\cite{Green:1997di}, the continuous symmetry is believed to be
broken down to discrete $E(7,7;Z)$ by non-perturbative effects
like instantons \cite{Bianchi:1998nk} and by
Dirac-Schwinger-Zwanziger charge quantization for regular black
holes. The breaking of $E(7,7;R)$ to $E(7,7;Z)$ is instrumental to
setting a uniform mass-gap, given by Planck Mass $M_{Pl}=
\sqrt{\hbar c/G_N} = 1.22\times 10^{19} GeV/c^2$, everywhere
inside the moduli space $E_{7(7)}/SU(8)$, for all regular BPS
states with $\cI_4\neq 0$ where \cite{Kallosh:1996uy} \be \cI_4 =
q^{abcd} \cQ_a \cQ_b \cQ_c \cQ_d \label{quartic} \ee is the
quartic Cartan invariant of $E_{7(7)}$ \cite{Cremmer:1979up} and
$\cQ_a$ is a 56-dimensional vector of `bare'
quantized\footnote{Perturbative Feynman graphs would not show
pathologies even in a non-anomalous but possibly chiral abelian
theory with charges in irrational ratios. Charge quantization
would however result from an embedding in a non-abelian theory.}
electric and magnetic charges with respect to the 28 $U(1)$ gauge
groups.

$\cN =8$ supergravity can be derived by dimensional reduction of
$\cN =1$ supergravity in $D=11$ on a 7-torus $T^7$ or of $\cN =2$
supergravity in $D=10$ on a 6-torus $T^6$. In such a process an
infinite number of states which would be massive in $D=4$ are
eliminated from the theory. After taking into account charge
quantization and assuming that the perturbative theory be UV
finite to all orders, we would like to propose the {\it
plausibility of a non-perturbative completion of genuinely
4-dimensional $\cN=8$ supergravity, that only include regular BH
states with $\cI_4 \neq 0$ and exclude all singular states with
$\cI_4 = 0$.}

Our proposal has some analogy with the story for pure $\cN=4$ SYM
in $D=4$, decoupled from gravity and other stringy interactions.
Even after including non-perturbative effects,  $\cN=4$ SYM in
$D=4$ should not be thought of as a compactification of Type I or
Heterotic strings, that contain the same massless states but
differ by the massive completion, but rather in terms of the
AdS/CFT correspondence \cite{Maldacena:1997re}. Pure 4-dimensional
$\cN=8$ supergravity, including regular non-perturbative states,
may be disconnected from toroidal compactifications of Type II
superstrings, that unavoidably give rise to 1/2 BPS states with
$\cI_4=0$. The fact that all known $\cN=8$ supergravity
perturbative amplitudes could be expressed in terms of $\cN=4$ SYM
amplitudes in the superconformal phase, where the latter enjoys 32
supersymmetries, 16 of Poincar\'e type and 16 superconformal,
might be more than an analogy in this respect.

The conjectured UV finiteness of $\cN =8$ supergravity, associated
with continuous $E_{7(7)}$ symmetry, has been questioned by Green,
Ooguri, Schwarz  in \cite{Green:2007zzb}, where non-decoupling of
BPS states from four-dimensional $\cN =8$ supergravity was
discussed. The main conclusion of \cite{Green:2007zzb} was that
the $\cN =8$ supergravity limit of string theory does not exist in
four dimensions, irrespective of whether or not the perturbative
approximation is free of UV divergences. String theory adds to the
256 massless states of four-dimensional $\cN =8$ supergravity an
infinite tower of states, such as Kaluza-Klein momenta and
monopoles, wound strings and wrapped branes.

Later on we will study these states as nearly massless extremal
black hole solutions of non-perturbative four-dimensional $\cN =8$
supergravity.

Classical solutions of the $\cN =8$ version of non-linear Einstein
equations include stable, zero temperature, extremal, BPS and
non-BPS charged black holes \cite{Ferrara:1997tw}.  For
appropriate choices of the charges, these can be viewed as smooth
solitons interpolating between flat Minkowski space-time at
infinity and Bertotti-Robinson $AdS_2 \times S^2$ geometry near
the horizon. The asymptotic values of the scalar fields are
largely arbitrary and determine the ADM mass $M$ for given charges
$\cQ_a$. Thanks to the attractor mechanism, their near-horizon
values are determined in terms of the charges
\cite{Ferrara:1995ih}. The entropy of a black hole in $\cN =8$
supergravity   is related to the horizon area by the
Bekenstein-Hawking formula\footnote{In string theory, where the
entropy is computed via degeneracy of states, formula
(\ref{quartic}) is believed to be valid only for large charges. It
is believed that 1/2 BPS states have zero entropy even at the
quantum level. For 1/4 BPS states the situation is still somewhat
puzzling \cite{Sen:2009bm}.}  \be S_{BH} = {1\over 4} A_H = \pi
\sqrt{|\cI_4|} \ .  \ee $\cN =8$ attractors were studied in
\cite{Ferrara:2006em}.

\

{\it The  ADM mass of extremal black holes} -- The ADM mass of an
$\cN =8$  extremal  black hole depends on its charges and on the
asymptotic values of the scalar fields, both  transforming under
$E_{7(7)}$ symmetry: $M_{BH} = M_{ADM}(\cQ,\phi)$. A manifestly
$E_{7(7)}$ covariant expression for the mass is related to the
maximal eigenvalue of the central charge matrix
 \cite{Hull:1994ys},\cite{Ceresole:1995jg}
 \be M_{ADM}^2(\cQ,\phi) \geq {\rm
Max}_i\{|Z_i(\cQ,\phi)|^2\} \ee where $Z_i(\cQ,\phi), i=1,2,3,4$,
are the four (skew) eigenvalues of the `dressed' central charge
matrix $Z_{AB}(\cQ,\phi) = \cQ_a \cV^a{}_{AB}(\phi) $ (with
$\bar{Z}^{AB}(\cQ,\phi) = \cQ_a \cV^{a,AB}(\phi) $), while $\cQ_a
$ are the `bare' quantized charges. The coset representative for
$E_{7(7)}/SU(8)$ can be taken to be of the form: \be
\cV^{a,\hat{a}}(\phi)=\{\cV^a{}_{AB}(\phi);\cV^{a,AB}(\phi)\}
\label{cosetrep}\ee with $a,\hat{a}=1,...,56$ while $A,B=1,...8$
so that $[AB]$ runs over the ${\bf 28}$ and ${\bf 28^*}$ of
$SU(8)$ \cite{Cremmer:1979up}. The positive hermitian matrix $
\cH_A{}^B = Z_{AC}\bar{Z}^{BC} $ has four {\it real positive
eigenvalues} $|Z_i(\cQ,\phi)|^2=\lambda_i$, which can be put in
decreasing order $ \lambda_1\geq \lambda_2\geq \lambda_3 \geq
\lambda_4 $. The BPS condition requires that the ADM mass be
exactly equal to the largest eigenvalue of the central charge
matrix \be M_{ADM}^2(\cQ,\phi)_{BPS} = {\rm
Max}_i\{|Z_i(\cQ,\phi)|^2\} \ee Non-BPS extremal black holes have
a mass which is strictly larger than the largest eigenvalue of the
central charge matrix \be M_{ADM}^2(\cQ,\phi)_{non-BPS} > {\rm
Max}_i\{|Z_i(\cQ,\phi)|^2\} \ee In such a case the quartic
invariant is strictly negative, the non-BPS extremal black hole
geometry is regular and its mass is never zero,  $ \cI_4<0$,
$S_{BH}^{non-BPS}= {1\over 4} A_H = \pi \sqrt{-\cI_4} $.

 In terms
of the central charge matrix, the quartic invariant in the
area/entropy formula (\ref{quartic}) reads
 \be \cI_4 (\cQ,\phi) = Tr[(Z\bar{Z})^2] - {1\over 4} [Tr(Z\bar{Z})]^2 + 8 Re Pf(Z) \ee
It is interesting that each $SU(8)$ invariant term in this
expression depends on the moduli, but the total expression is
moduli independent due to $E_{7(7)}$ symmetry.

{ \it The bounds on the ADM mass and horizon area for  1/8, 1/4,
1/2 BPS  } -- The definition of the fraction of unbroken
supersymmetry (1/8, 1/4, 1/2) on BPS solutions of $\cN =8$
supergravity in terms of the properties of the eigenvalues of the
central charge is based on the classification of the orbits of
exceptional groups in \cite{Ferrara:1997uz}.  Here we present
 the bounds on the moduli dependent ADM mass as well
as the bounds on the horizon area:
%\begin{itemize}
%\item

1/8 BPS case: $\lambda_1> \lambda_2\geq \lambda_3
\geq\lambda_4\geq 0 $.  This is consistent with a non-negative
$\cI_4\geq 0$. The mass is constrained as follows
 \be
 {1\over 8} \Tr Z\bar{Z} < (M_{ADM}^2(\cQ,\phi))_{1/8}\leq {1\over 2} \Tr Z\bar{Z}
 \ee
% \item
1/4 BPS case: $\lambda_1= \lambda_2> \lambda_3 =\lambda_4\geq 0$.
 This is consistent with $\cI_4 = 0$ and
      \be
{1\over 8} \Tr Z\bar{Z} < (M_{ADM}^2(\cQ,\phi))_{1/4}\leq  {1\over
4} \Tr Z\bar{Z}
 \ee

1/2 BPS case: $\lambda_1= \lambda_2= \lambda_3 =\lambda_4\geq 0$.
 This is consistent with $\cI_4 = 0$ and
\be (M_{ADM}^2(\cQ,\phi))_{1/2}= {1\over 8} \Tr Z\bar{Z}
 \ee

 Thus, taking into account that $ {1\over 2} \Tr Z\bar{Z} =
\lambda_1+  \lambda_2+ \lambda_3+\lambda_4 $ is a {\it sum of
positive terms}, in all cases above the ADM mass of BPS states can
only vanish when
  \be
 \Tr \, Z\bar{Z}(\cQ,\phi)\equiv \cQ_a \cV^a{}_{AB}(\phi) \,
\cV^{b,AB}(\phi) \cQ_b \label{BPZ} \rightarrow 0
  \ee
with $ \cV^{a,\hat{a}}(\phi)$ a coset representative
(\ref{cosetrep}). It is a $56\times 56$ square matrix which is
invertible at any point inside the $E_{7(7)}/SU(8)$ moduli space
\cite{Cremmer:1979up}. Since the 56-charge vector $\cQ_a$  must
have some non-zero elements (for extremal black holes) {\it the
mass may vanish only at the boundary of moduli space}
\cite{Hull:1994ys}. At the boundary, by definition, the coset
representative degenerates, the perturbative theory is not valid,
it blows up. The boundary is at infinite distance from any
interior point as it corresponds to some scalar fields with
canonical kinetic terms $\sim (\partial \phi)^2$ going to
infinity, $\phi \rightarrow\pm \infty$.

Consider states which are massive  in the interior of moduli space
and can become massless at the boundary.

{\it All cases of $M_{ADM}$ reaching zero at the boundary} --
Starting from a scalar (spherically symmetric) state, the maximum
spin is: $J_{Max} = (\cN - K)/ 2$, where $K/\cN$ is the fraction
of susy preserved, $J_{Max} = 2,\, 3,\, 7/2, \, 4$ respectively
for 1/2 (ultrashort), 1/4 (short), 1/8 (minimally short) BPS and
non-BPS  (long) multiplets \cite{Ferrara:1980ra}.

{\it  1/8 BPS case}. In this case we have two classes of
solutions.

a) The first class, with $\cI_4 > 0$ and regular area of the
horizon, has a mass gap which due to the attractor mechanism
\cite{Ferrara:1995ih} defines the minimal value of the ADM mass
via the one at the horizon \be M^2_{ADM} \geq M^2_{hor}= \cI_4 >0
\ee The bound is saturated for double-extreme black holes with
constant scalars where $M^2_{ADM} = M^2_{hor}= \cI_4>0$.

b) The second class of 1/8 BPS states, with $\cI_4 = 0$, is
somewhat special. In the interior of  moduli space, the
eigenvalues of the central charge matrix are different,
$\lambda_1> \lambda_2\geq \lambda_3 \geq\lambda_4\geq 0 $ and the
mass is non-zero. However, at the boundary the mass vanishes. The
eigenvalues vanish at different rates: ${\Delta \sqrt \lambda/
\sqrt \lambda} \neq 0$. Thus in the interior of moduli space, 1/8
BPS solutions have a non-vanishing ADM mass and form multiplets
with maximal spin $J_{Max} = 7/2$ at least.

{\it  1/4 BPS case}, $\cI_4 = 0$ always. At the boundary of moduli space, the
mass vanishes. The ratio ${\Delta \sqrt \lambda/
\sqrt \lambda}$
 %of the splitting of the eigenvalues with respect to the eigenvalues
depends on the way the boundary is approached. In
some limit, both $\Delta \sqrt \lambda$ as well as $ \sqrt
\lambda$ vanish asymptotically. In some other limit, the ratio may
not vanish. This seems to indicate that, even at the boundary, 1/4
BPS states behave differently from 1/2 BPS states, whose masses
vanish faster towards the boundary.  In any case, all 1/4 BPS solutions
 are singular, both massive and massless.

{\it 1/2 BPS  case}. In the interior of moduli space, $\lambda_1=
\lambda_2= \lambda_3 =\lambda_4> 0$, while $\lambda_1=\lambda_2=
\lambda_3 =\lambda_4= 0$ at the boundary. This is only consistent
with $\cI_4 = 0$. As in the above cases, 1/2 BPS states are
massive in the interior and massless at the boundary: \bea
(M_{ADM}^2(\cQ,\phi\in{Interior}))_{1/2}= \lambda_i >0 \\
(M_{ADM}^2(\cQ,\phi\in{Boundary}))_{1/2}= \lambda_i =0
 \eea
All 1/2 BPS solutions are singular since the area of the horizon
vanishes. Of all BPS solutions so far discussed, the infinite
tower of 1/2 BPS states is most easily identified with the states
described in \cite{Green:2007zzb}. They represent the K-K and
string states obtained by compactification down to four dimensions
in addition to the massless 256 states of $\cN =8$ supergravity.
When viewed as non-perturbative solution of Einstein equations
they represent singular space-times:  {\it For all solutions with
$\cI_4 = 0$ the area of the horizon vanishes, even when the
solutions are massive in the interior of moduli space}. Although
the attractor mechanism does not work for singular solutions, the
algebraic feature of $\cN =8$ supergravity that $\cI_4$ is moduli
independent  gives us complete control on the situation.  The
emergence of massless higher spins at the boundary for 1/4 and 1/8
BPS multiplets with $\cI_4 =0$ corresponds to multi-charge
solitonic states with finite entropy in higher dimensions.

In comparing $D=4$ $\cN =8$ supergravity with string or M theory
one should keep in mind that the 70 scalars of $ {E_{7(7)}/SU(8)}$
include on an equal footing all massless moduli which arise in
dimensional reductions from higher dimension. From a superstring
perspective, they also contain the string coupling $g_s$, so that
the boundary of moduli space may encompass many limits of string
theory, such as infinite string coupling, zero string tension,
de-compactifications and others \cite{Witten:1995ex}. Moreover the
ADM mass should be considered in the Einstein frame, where
Newton's law holds, and not in the string frame. $E_{7(7)}$
invariance of the $D=4$ space-time geometry is only manifest in
the Einstein frame. In the string frame, the metric itself would
transform and $M_s = g_s M_{Pl}$ would vary as $g_s$ at fixed
$M_{Pl}$.

{\it Space-time geometry} --  Regular extremal black hole
solutions have the following infinite throat  geometry
\cite{Ferrara:1997tw}: \be ds^2= -e^{2U(\tau)} dt^2+ e^{-2U(\tau)}
\left ({d\tau^2\over \tau^4}+ {1\over \tau^2} d^2\Omega \right)
\ee The ADM mass defines the behavior of the metric at asymptotic
infinity in the space-time at $\tau \rightarrow 0$ \be U(\tau
)_{\tau \rightarrow 0} = M_{ADM}(\cQ,\phi_{r\rightarrow \infty})\,
\tau \ee Near the regular horizon at $\tau \rightarrow -\infty $,
where the geometry approaches $AdS_2 \times S^2$, one finds \be
e^{-2U(\tau)}_{\tau \rightarrow-\infty} \rightarrow
\sqrt{|\cI_4(\cQ)|} \tau^2 \ee As it is clear from the metric, the
area of the horizon equals $4\pi \sqrt{|\cI_4(\cQ)|}$. The
attractor mechanism in this case requires that $M_{ADM}^2
(\cQ,\phi_{r\rightarrow \infty})\geq \sqrt{|\cI_4(\cQ)|}$. Only
for configurations with $\cI_4(\cQ)=0$ the ADM mass may vanish:
since the area is moduli independent, once the mass vanishes at
the  boundary, which requires $\cI_4(\cQ)=0$, the area vanishes
also for all massive solutions with the corresponding set of
charges. {\it  In all such cases the null surface of the horizon
turns into a null singularity}.\footnote{The singularity of the
1/2 BPS states in $D=4$ space-time was not considered pathological
because in the string frame the so-called $\sqrt 3$ dilatonic
black hole has a regular horizon. However physical ADM masses are
defined in the Einstein frame, where such a black hole is
singular.}

In order to consider including states with $\cI_4 =0$ in the
non-perturbative $\cN =8$ supergravity spectrum, one should find a
way to resolve their singularity in $D=4$. The higher dimensional
resolution of dilatonic black hole singularities is well
understood \cite{Gibbons:1994vm}. However, the dilemma we face
here is to study the issue directly in $D = 4$ in view of the
possible UV finiteness of $N = 8$ supergravity. In higher
dimensions maximal supergravity is expected to be perturbatively
divergent at $L$ loops when $ D < 4 + 6/L $ is violated
\cite{Green:2006yu, Bern:2009kd}, and to be consistent only as
part of a more fundamental theory such as string or M- theory.
Without invoking the latter, we can only rely on known methods in
$D=4$.

{\it Resolution of singularity by non-extremality?} One can try to
regularize the singular solutions by small deviation from
extremality\footnote{This was suggested by G.~Horowitz and
S.~Shenker.}.  A set of regular non-extremal charged black hole
solution with non-vanishing area of the outer horizon are
presented in eq.(23) of \cite{Kallosh:1992ii}. As one approaches
the extremal limit along the ``large orbit'' \cite{Ferrara:1997uz}
$\cI_4(\cQ)= 2|PQ|\neq 0$ where
 the area is regular,  the non-extremal black holes have
zero limiting temperature and remain regular. These are always
massive. On the other hand, along the ``small orbit'' with
${\cI_4(\cQ)}= 2|PQ|= 0$, the Hawking temperature  approaches a
non-zero constant as shown in eqs. (64), (65) in
\cite{Kallosh:1992ii}. Meanwhile, the so called $\sqrt 3$
dilatonic black holes, which are regular only when non-extremal,
in the limit to extremality have a diverging Hawking temperature
\cite{Dowker:1994up}. The discontinuity and divergence of
temperature in the extremal limit suggest the breakdown of the
thermodynamic description. No alternative descriptions are known
in $\cN =8$, as we show below.

{\it Resolution of singularity by stretching the horizon?} For
Calabi-Yau compactifications of string theory with non-zero second
Chern class, there  are known quantum corrections which
 stretch
the  horizon area of ``small black holes'', for instance, in IIB
string theory compactified on $K3\times T^2$
\cite{Dabholkar:2004dq}.   In $\cN =8$ supergravity there are no
known quadratic or higher curvature corrections  which would be
able to stretch the horizon and remove the null singularity of
nearly massless and massless ``black holes''. In fact if the
theory were perturbatively UV finite no corrections whatsoever
would appear to the lowest two-derivative action.

{\it Resolution of singularity by the enhan{c}on mechanism?} The
enhan{c}on mechanism \cite{Johnson:1999qt} is based on a special
property of extended supergravities with $\cN = 3,4 $. They admit
matter vectors with their associated `matter' central charges. The
two-derivative action is completely fixed by the number of vector
multiplets and the structure constants of the gauge algebra. The
theory retains supersymmetry when the number of matter multiplets
changes. Therefore supersymmetry does not prevent transitions from
one version of the theory to another, with different number of
matter multiplets.  {\it In $\cN <5 $ supergravities generically
the matrix relating the central charges $Z_{AB}$ to `bare' charges
$\cQ_a$ is a rectangular rather than a square matrix, due to the
presence of the matter multiplets}  \cite{Ceresole:1995jg},
\cite{Hull:1994ys}. {\it Therefore 1/2 BPS states can become
massless in the interior of moduli space}. Typically, these points
of enhanced gauge symmetry coincide with orbifold points which can
be seen in perturbative string theory. This phenomenon is the
basis of the so-called enhan{c}on mechanism which resolves the
singularity of repulson solutions in supergravity. An analogous
phenomenon takes place in $\cN =2$ theories where the number of
matter multiplets can change at finite distance in moduli space
\cite{Strominger:1995cz}, \ie at points where a BPS state becomes
massless due to quantum corrections to the prepotential of the
form $ \cF_{1-loop}^{\cN=2}  \sim {1\over 2\pi} Z^1 \log Z^1 $.

The two derivative action of  $\cN = 8$ supergravity is unique and
it describes one single supermultiplet (the same is also true for
$\cN =5,6$).  As a result, {\it central charges are related to the
`bare' charges by a square invertible matrix, since $\cN \ge 5$
supergravities do not admit matter multiplets. Therefore new
massless BPS states cannot appear in the interior of moduli space
and  cannot support enhan{c}on type transitions.} Thus, no
enhan{c}on resolution of singularities  applies in $\cN=8$
supergravity, viewed as a pure four-dimensional theory.

Since single-center states with $\cI_4 = 0$ are singular,
irrespectively of any BPS condition, we expect that the lowest
mass states in this sector be a BPS multi-center black hole, \ie a
bound state of regular BPS states with $\cI_4\neq 0$. The mass
scale is set by the unique mass scale of theory, Planck mass,
while the relative positions are constrained by the BPS condition
in terms of the $E(7,7)$ invariant pairings of the constituents'
charges. Since singular single-center BH's with $\cI_4 = 0$ cannot
be formed because of the weak cosmic censorship hypothesis, two or
more regular BH's whose total charge gives $\cI_4=0$, may only
form a multi-center regular bound state rather than a
single-center singular object.

We conclude with remarks on the conjectured UV finiteness of
perturbative $\cN =8$ $D=4$ supergravity \cite{Bern:2009kd} and
 massless string states described in
\cite{Green:2007zzb}. If the perturbative theory has some, yet
unknown  UV divergences \cite{Green:2010sp, Bjornsson:2010wm}, our
analysis of the massless black hole solutions may require
modifications. However, if there are no UV divergences in
perturbation theory, we do not expect corrections to our analysis
of the space-time properties of these states. Here we have shown
that the states in \cite{Green:2007zzb} are singular. There are
two logical possibilities. 1. These states may be consistently
excluded from the four-dimensional theory and therefore do not
affect UV properties of $\cN =8$ $D=4$ supergravity 2. They can be
proven to be required in $D=4$ and affect the perturbative theory.
Here we have given circumstantial evidence for the plausibility of
the first option. Future investigations may shed further light on
this.

\

We are grateful to T. Banks, Z. Bern, L. Dixon, M. Green, M.
Gunaydin, G. Horowitz, A. Linde, D. Simic, S.~Shenker, R. Stora, A.
Strominger,   L. Susskind and V. S. Varadarajan,  for most stimulating discussions.
This work was partially supported by the ERC Advanced Grant
n.226455 {\it ``Superfields''} and by the Italian MIUR-PRIN
contract 2007-5ATT78 {\it ``Symmetries of the Universe and of the
Fundamental Interactions''}. The work of SF has been supported in
part by DOE grant DE-FG03-91ER40662, Task C. The work of RK  is
supported by the NSF grant 0756174.

\end{document}